\documentclass[a4paper,11pt]{article}

\usepackage{jcappub} 

\usepackage[T1]{fontenc} 

\usepackage{amssymb}
\usepackage{epsfig} 
\usepackage{amsmath} 
\usepackage{graphicx}

\usepackage{tikz}
\usetikzlibrary{positioning,arrows}
\usetikzlibrary{decorations.pathmorphing}
\usetikzlibrary{decorations.markings}

\newcommand{\beq}{\begin{equation}}
\newcommand{\eeq}{\end{equation}}
\newcommand{\bea}{\begin{eqnarray}}
\newcommand{\eea}{\end{eqnarray}}
\newcommand{\ba}{\begin{array}}
\newcommand{\ea}{\end{array}}
\newcommand{\bi}{\begin{itemize}}
\newcommand{\ei}{\end{itemize}}
\newcommand{\bn}{\begin{enumerate}}
\newcommand{\en}{\end{enumerate}}
\newcommand{\bc}{\begin{center}}
\newcommand{\ec}{\end{center}}
\renewcommand{\l}{\left}
\renewcommand{\r}{\right}

\newcommand{\eq}[1]{Eq.~(\ref{#1})}
\newcommand{\eqs}[2]{Eqs.~(\ref{#1}) and (\ref{#2})}

\newcommand{\MeV}{\mathinner{\mathrm{MeV}}}
\newcommand{\GeV}{\mathinner{\mathrm{GeV}}}

\title{\boldmath Higgs portal vector dark matter for $\GeV$ scale $\gamma$-ray 
excess from galactic center}

\author{P. Ko, Wan-Il Park and Yong Tang}

\affiliation{School of Physics, KIAS, Seoul 130-722, Korea}

\emailAdd{pko@kias.re.kr}
\emailAdd{wipark@kias.re.kr}
\emailAdd{ytang@kias.re.kr}

\abstract{
We show that the $\GeV$ scale $\gamma$-ray excess from the direction of the Galactic 
Center can be naturally explained by the pair annihilation of Abelian vector dark 
matter (VDM) into a pair of dark Higgs bosons $VV\rightarrow \phi \phi$, followed 
by the subsequent decay of $\phi$ into $\phi \rightarrow b\bar{b} , \tau \bar{\tau} $.  
All the processes are  described by a renormalizable  VDM model with the Higgs portal, which is naturally flavor-dependent. 
Some parameter space of this scenario can be tested at the near future direct dark 
matter search experiments  such as LUX and XENON1T.
}

\begin{document}
\maketitle
\flushbottom

\section{Introduction}
It has been known for sometime that there is anomalous $\GeV$ scale $\gamma$-ray 
excess from the direction of the Galactic Center \cite{Goodenough:2009gk,Hooper:2010mq,Boyarsky:2010dr,Hooper:2011ti,Abazajian:2012pn,Gordon:2013vta,Hooper:2013rwa,Huang:2013pda,Abazajian:2014fta,Daylan:2014rsa}.
Interestingly, the excess seems to be remarkably well described by an expected signal 
from 31-40 $\GeV$ dark matter (DM) annihilating dominantly to $b\bar{b}$ with a cross 
section $\sigma v \simeq \l(1.7-2.3\r) \times 10^{-26} {\rm cm}^3/{\rm s}$ \cite{Daylan:2014rsa,Steigman:2012nb}. (See also Ref.~\cite{Yuan:2014rca} for millisecond pulsars as an astrophysical alternative.)  
Because of the importance of DM pair annihilation into $b\bar{b}$ for the GC $\gamma$-ray excess, some ideas on flavored DM have been 
put forth~\cite{Berlin:2014tja,Agrawal:2014una}. 

We should note that it is the shape of $\gamma$ spectrum from dark matter annihilation that mainly matters rather than the precise value for $\sigma v$ since there is a large uncertainty in the density profile of dark matter near the Milky Way center. As long as $\langle \sigma v\rangle \left( \rho_{\textrm{DM}}/M_\textrm{DM} \right)^2$ is at the right amount, a good 
fit can be achieved for $b\bar{b}$ final state. Actually, $b\bar{b}$ does not need to be 
the only annihilation channel, it was shown~\cite{Daylan:2014rsa} that flavor-dependent annihilations can also fit the data well. 
Such kind of flavor-dependent annihilations may indicate a Higgs-like scalar mediator,
since Higgs-like scalar will couple with the heaviest particle it can couple to.

The required cross section is very close to the canonical value for neutral thermal relic 
dark matter.   It can be achieved either $s$-wave annihilation or $p$-wave annihilation 
with $s$-channel resonance at present.  However, in the latter case, the resonance band 
is likely to be very narrow that leads to a severe fine-tuning, which is not that attractive.
With this consideration, perhaps the simplest scenario for dark matter model that 
can explain the $\gamma$-ray excess would be those involving scalar mediator with 
Higgs portal interaction(s), since in this case the scalar mediator will couple strongly 
to the $b\bar{b}$, the heaviest particles kinematically producible 
\footnote{Another possibility would be to consider flavored DM
~\cite{Berlin:2014tja,Agrawal:2014una}. }.
Then,  one can imagine the following simple scenarios of DM having $s$-wave 
annihilation channel:
\begin{enumerate}
\item Singlet scalar dark matter (SSDM): a real scalar mediator \cite{Modak:2013jya}
\item Singlet fermion dark matter (SFDM): a pseudo-scalar mediator \cite{Boehm:2014hva,Izaguirre:2014vva,Ipek:2014gua}
\footnote{While we were working on these possibilities, this paper was put on the archive, 
and we don't consider this possibility any more in this work.}
\item Singlet vector dark matter (SVDM): a real scalar mediator \cite{Boehm:2014bia,Baek:2012se,Baek:2014goa}
\end{enumerate}
Note that the structure of above scenarios can be realized easily when DM is charged 
under a dark gauge symmetry which is broken to, for example, a discrete $Z_2$ or 
$Z_3$ symmetry.   Hence those scenarios would also work equally well.
For other recent proposals of DM models to address the $\GeV$ $\gamma$-ray spectrum, see Refs. \cite{Okada:2013bna,Alves:2014yha,Berlin:2014tja,Agrawal:2014una}.

Potentially the most important constraint on those singlet dark matter models may 
come from  direct search experiments, for example, LUX \cite{Akerib:2013tjd}.
However the existence of extra scalar boson mediating dark and visible sectors via Higgs portal interaction(s) has a significant effect on direct searches if the mass of the extra 
non-SM Higgs is not very different from that of SM Higgs \cite{Baek:2011aa}, and the 
constraint from direct searches can be satisfied rather easily.  Note that this feature is not 
captured at all in effective field theory approach, and it is important to work on the 
minimal renormalizable and unitary Lagrangian for physically sensible results
\footnote{See Refs.~\cite{Baek:2011aa,Baek:2012se} for the original discussions 
on this point, and Ref.~\cite{HiggsPortalEFT}  for more discussion on the correlation 
between the invisible Higgs branching ratio and the direct detection cross section in the 
Higgs portal SFDM and SVDM models,}.

In this paper, we revisit SVDM scenario with Higgs portal in the context of the the 
$\gamma$-ray excess from the Galactic Center, and show that the SVDM model can 
naturally explain it, 
while satisfying all of known constraints coming from CMB, Fermi-LAT $\gamma$-ray 
search and LHC experiments.  We also show that the parameter space relevant for the 
$\gamma$-ray excess can be probed by the near future direct dark matter search 
experiment, for example LUX and XENON1T.

This paper is organized as follows.
In Section~\ref{sec:model}, we recapitulate the renormalizable SVDM model with 
Higgs portal.   In Section~\ref{sec:const}, various  relevant constraints on the model 
are discussed, including relic density estimation, vacuum stability, collider bounds, 
CMB and direct detection cross section, etc., and we show that our model can explain 
the $\gamma$-ray excess from the galactic center without any conflict with other 
cosmological and astrophysical observations.   
In Section~\ref{sec:conc}, our conclusion is drawn.

\section{The renormalizable SVDM with Higgs portal} \label{sec:model}

Let us consider a Abelian vector boson dark matter 
\footnote{
The Abelian VDM was first considered in Ref.~\cite{Lebedev:2011iq} where 
the VDM mass assumed to be generated either by the St\"{u}ckelberg  or 
by dark Higgs mechanism, but the role of dark Higgs boson was ignored 
within effective field theory (EFT). 
However, in the presence of the dark Higgs boson, the resulting VDM phenomenology can 
be vastly different from the one in the VDM model of EFT. 
See Ref.~\cite{Baek:2012se} for more detailed discussion.}, 
$X_\mu$, which is assumed to be
a gauge boson associated with Abelian dark gauge symmetry $U(1)_X$. 
The simplest model will be defined with a complex scalar dark Higgs field $\Phi$ only, 
and no other extra fields.  The VEV of $\Phi$ breaks $U(1)_X$ spontaneously and 
generate the mass for $X_\mu$ through the standard Higgs mechanism 
 (see also Ref.~\cite{Farzan:2012hh}):
\begin{eqnarray}
{\cal L}_{VDM} & = & - \frac{1}{4} X_{\mu\nu} X^{\mu\nu} +  
(D_\mu \Phi)^\dagger (D^\mu \Phi) 
- \lambda_\Phi \l( \Phi^\dagger  \Phi - \frac{v_\Phi^2}{2} \r)^2
\nonumber \\
& & - \lambda_{\Phi H} \l(\Phi^\dagger \Phi - \frac{v_\Phi^2}{2}\r) \l(H^\dagger H - \frac{v_H^2}{2}\r) \ ,
\label{eq:full_theory}
\end{eqnarray}
in addition to the SM Lagrangian which includes the Higgs potential term
\begin{equation}
\Delta {\cal L}_{\rm SM} = 
- \lambda_H \l( H^\dagger  H - \frac{v_H^2}{2} \r)^2.
\end{equation}
The covariant derivative is defined
as 
\[
D_\mu \Phi = (\partial_\mu + i g_X Q_\Phi X_\mu) \Phi ,
\]
where $Q_\Phi \equiv Q_X(\Phi)$ is the $U(1)_X$ charge of $\Phi$ and we will take $Q_\Phi=1$ throughout
the paper.

Assuming that the $U(1)_X$-charged complex scalar $\Phi$ develops a nonzero VEV, 
$v_\Phi$, and thus breaks $U(1)_X$ spontaneously, we would have 
\[
 \Phi  = \frac{1}{\sqrt{2}} \left( v_\Phi+ \phi(x) \right) .
\]
Therefore the Abelian vector boson $X_\mu$ gets mass $M_X = g_X |Q_\Phi| v_\Phi$. 
And the hidden sector Higgs field (or dark Higgs field) $\phi (x)$ will mix with 
the SM Higgs field $h(x)$ through the Higgs portal  $\lambda_{\Phi H}$ term, resulting 
in two neutral Higgs-like scalar bosons.
The mixing matrix $O$ between the two scalar fields is defined as
\begin{equation}
\left(
 \begin{array}{c}
  h \\ \phi
 \end{array}
\right)
= O 
\left(
 \begin{array}{c}
  H_2 \\ H_1
 \end{array}
\right)
\equiv  
\left(
 \begin{array}{cc}
  c_\alpha & s_\alpha \\
  -s_\alpha & c_\alpha 
 \end{array}
\right)
\left(
 \begin{array}{c}
  H_2 \\ H_1
 \end{array}
\right),
\end{equation}
where $s_\alpha (c_\alpha) \equiv \sin\alpha (\cos\alpha)$, $h, \phi$ 
are the interaction eigenstates and $H_i (i=1,2)$ are the
mass eigenstates with masses $m_i$, respectively. 
The mass matrix in the basis $(h,\phi)$ can be written 
in terms either of Lagrangian parameters or of the physical parameters as follows:
\begin{equation}
\left(
\begin{array}{cc}
2 \lambda_H v_H^2 & \lambda_{\Phi H} v_\Phi v_H \\ 
\lambda_{\Phi H} v_\Phi v_H & 2 \lambda_\Phi v_\Phi^2
\end{array}
\right) =
\left(
\begin{array}{cc}
m_1^2 s_\alpha^2 + m_2^2 c_\alpha^2 & (m_2^2 -m_1^2 )s_\alpha c_\alpha \\ 
(m_2^2 -m_1^2 )s_\alpha c_\alpha & m_1^2 c_\alpha^2 + m_2^2 s_\alpha^2
\end{array}
\right).
\label{eq:mass_matrix}
\end{equation}
Note that one can take $m_1 , m_2$ and $\alpha$ are independent parameters.

In the small mixing limit which is of our interest, the mass eigenstates are approximated to the interaction eigenstates as $\l( H_2, H_1 \r) \approx \l( h, \phi \r)$, and we use $\l( h, \phi \r)$ to represent quantities associated with $\l( H_2, H_1 \r)$ from now on.

\section{Constraints} \label{sec:const}
Our VDM interacts with SM sector via Higgs portal interaction.
This means that it is subject to constraints from CMB observations, direct/indirect DM searches, and collider experiments.
However, for $30 \GeV \lesssim m_V \lesssim 80 \GeV$, constraints from CMB \cite{Madhavacheril:2013cna} and indirect searches \cite{Hooper:2012sr,Gordon:2013vta,Ng:2013xha,Kong:2014haa} can be easily satisfied in our scenario as far as there is no enhancement of annihilation rate relative to the one at freeze-out.
So, in this section we consider only low energy phenomenology, direct detection and relic density.

\subsection{Vacuum stability}
The mixing between Higgs fields ($H$ and $\Phi$) causes a tree-level shift of 
$\lambda_H$ relative to that of SM in such a way that the relation 
\beq
\lambda_H = \l[ 1 - \l( 1 - \frac{m_\phi^2}{m_h^2} \r) \sin^2 \alpha \r] \frac{m_h^2}{2 v_H^2}
\eeq 
holds.  Hence, for $m_\phi < m_h$ one obtains $\lambda_H$ even smaller than that of 
SM, and vacuum instability of SM Higgs potential becomes worse.
So, it is better to take $\alpha$ as small as possible.   

Although tree-level mixing does not work, vacuum instability can be improved by the additional contribution of $\lambda_{\Phi H}$ to the $\beta$-function of $\lambda_H$,
\[
\Delta \beta_{\lambda_H}=\frac{1}{16\pi^2}\lambda^2_{\Phi H} ,
\] 
For $\alpha \lesssim m_\phi/m_h$, one finds $\lambda_\Phi \approx g_X^2/2$ which should be $\mathcal{O}(10^{-2})$ as shown in Section~\ref{subsec:relic}.
Then, the tachyon-free condition, $\lambda_{\Phi H} < 2 \sqrt{\lambda_\Phi \lambda_H}$, results in $\lambda_{\Phi H} \lesssim 0.07$ for $\alpha$ and $m_\phi$ in the range of our interest.
It might be large enough to improve the vacuum stability.
The exact lower bound on $\lambda_{\Phi H}$ that stabilizes the EW vacuum up to Planck scale depends on the precise values of top quark mass and the strong coupling constant, the detailed discussion of which is beyond this paper.

\subsection{Collider bound}
For $m_V < m_h/2$, the SM Higgs boson can decay into two VDM which is invisible.
Recent analysis from collider experiments showed that the branching fraction of the Higgs
boson  into invisible particles should be constrained as \cite{Chatrchyan:2014tja}
\beq \label{BrInvLHCbnd}
{\rm Br}_h^{\rm inv} < 0.51
\eeq
However the bound was extracted for a effective-field-theoretic (EFT) VDM model.
In a renormalizable complete theory like the one we are considering, more parameters 
are involved than EFT model.  Hence, instead of \eq{BrInvLHCbnd}, we use
\beq \label{BrNonSMLHCbnd}
c_\alpha > 0.904 + {\rm Br}_h^{\rm non-SM}/2
\eeq
which is an approximation obtained from the result of Ref.~\cite{Chpoi:2013wga}, and ${\rm Br}_h^{\rm non-SM}$ is the branching fractions of the Higgs decay to DMs and non-SM Higgs.
In our SVDM scenario, ${\rm Br}_h^{\rm non-SM}$ is given by
\beq \label{BrNonSM}
{\rm Br}_h^{\rm non-SM} = \frac{s_\alpha^2 \Gamma_h^{\rm inv} + \Gamma_h^{\phi\phi}}{c_\alpha^2 \Gamma_h^{\rm SM} + s_\alpha^2 \Gamma_h^{\rm inv} + \Gamma_h^{\phi\phi}}
\eeq
where 
\bea
\Gamma_h^{\rm SM} &\simeq& 4.07 \MeV
\\
\Gamma_h^{\rm inv} &=& \frac{g_X^2}{8 \pi} \frac{m_V^2}{m_h} \l[ 1 + \frac{1}{2} \l( 1 - \frac{m_h^2}{2 m_V^2} \r)^2 \r] \l( 1 - \frac{4 m_V^2}{m_h^2} \r)^{1/2}
\\
\Gamma_h^{\phi \phi} &=& \frac{1}{32 \pi m_h} \lambda_{h\phi\phi}^2 \l( 1 - \frac{4 m_\phi^2}{m_h^2} \r)^{1/2}
\eea
with
\bea
\lambda_{h\phi\phi} 
&=& \lambda_{\Phi H} v_H c_\alpha^3 + 2 \l( 3 \lambda_H - \lambda_{\Phi H} \r) v_H c_\alpha s_\alpha^2 - 2 \l[ 3 \l( \lambda_\Phi - \lambda_{\Phi H} \r) v_\Phi \r] c_\alpha^2 s_\alpha - \lambda_{\Phi H} v_\Phi s_\alpha^3
\nonumber \\
&\sim& \lambda_{\Phi H} v_H c_\alpha^3 \simeq \frac{s_\alpha c_\alpha^4 \l( m_h^2 - m_\phi^2 \r)}{v_\Phi}
\eea
In the second line of the above equation, we assumed the first term dominates over the others in the small mixing limit.

Using \eqs{BrNonSMLHCbnd}{BrNonSM}, we can constrain the allowed ranges of 
$g_X$ and $\alpha$ as shown in the white region of the left-panel of 
Fig.~\ref{fig:alpha-gx-LUXbnd}.
Note that in Fig.~\ref{fig:alpha-gx-LUXbnd}, the mixing angle is constrained to be 
$\alpha \lesssim 7 \times 10^{-2}$ for $m_\phi = 60 \GeV$.
The the upper-bound of $\alpha$ is lowered down for a lighter $m_\phi$.
Note that the current LHC, LUX or the future XENON1T experiments cover only 
a part of the allowed parameter region in $( \alpha , g_X)$.  There is ample region of 
parameter space which cannot be explored directly  in any experiments.

\subsection{Direct detection}

\begin{figure}[h]
\centering
\includegraphics[width=0.45\textwidth]{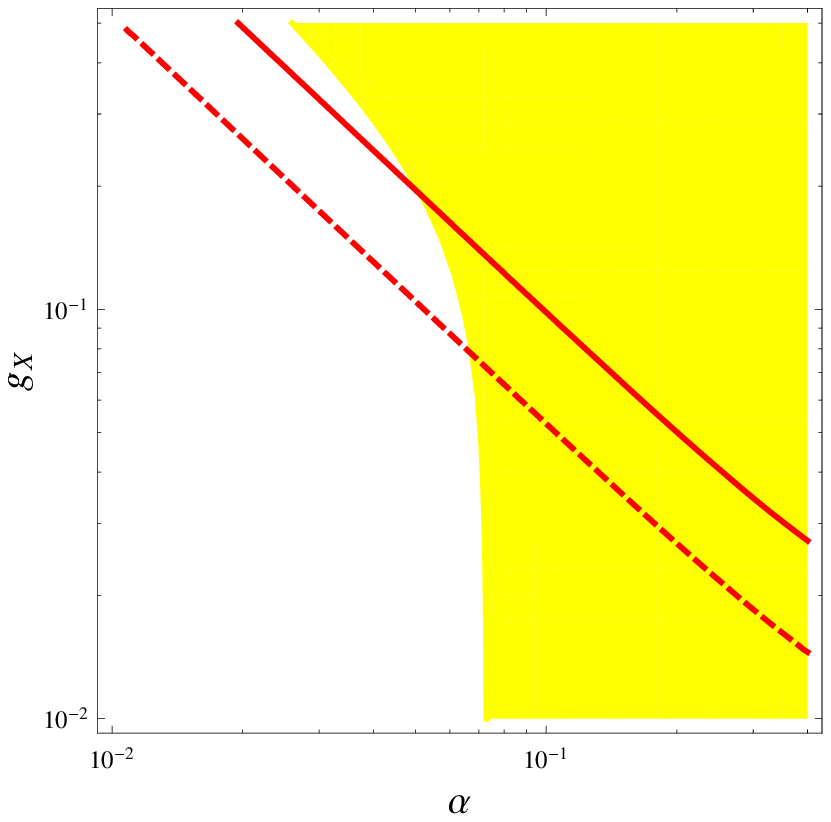}
\includegraphics[width=0.45\textwidth]{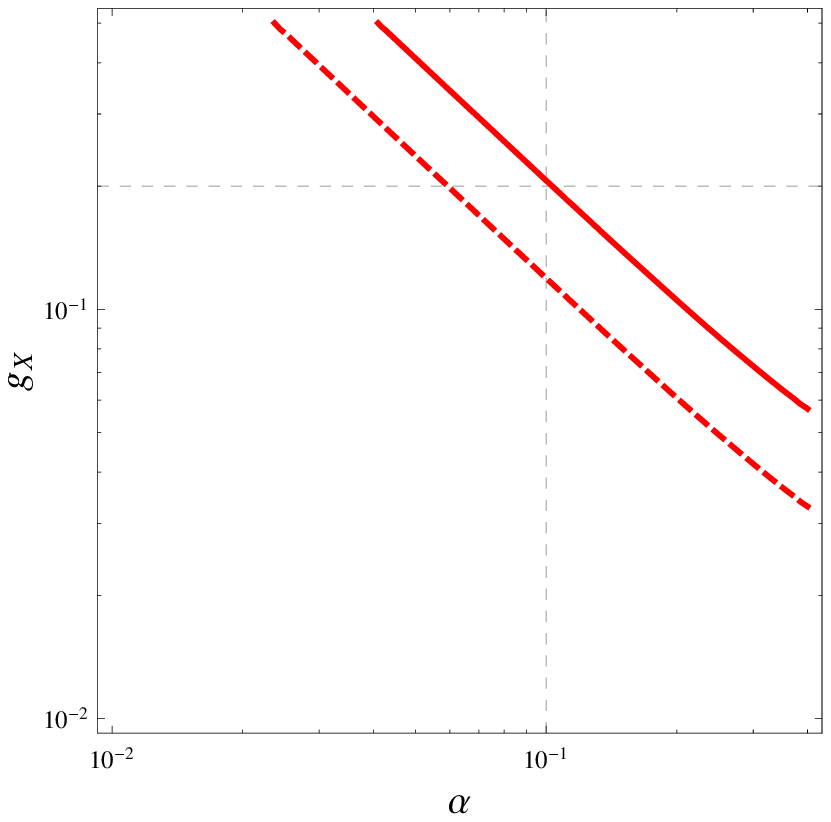}
\caption{\label{fig:alpha-gx-LUXbnd} 
A bound of $\alpha$ and $g_X$ from collider experiments, LUX and projected XENON1T direct DM searches \cite{Akerib:2013tjd} for $m_h=125$.
Left: $m_V = 35 \GeV$, $m_\phi = 60 \GeV$, and $\lambda_\Phi=0.1$.
Right: $m_V = 80 \GeV$, $m_\phi = 75 \GeV$.
Yellow region is excluded by collider constraint on ${\rm Br}_h^{\rm non-SM}$.
Solid and dashed red lines are upper-bound of DM-nucleon scattering cross section from LUX and XENON1T, respectively.
}
\end{figure}

For $30 \GeV \lesssim m_V \lesssim 80 \GeV$, LUX  experiment  for direct detection
of WIMP imposes a strong upper bound on the spin-independent (SI) dark matter-proton scattering cross section~\cite{Akerib:2013tjd} as:
\beq 
\sigma_p^{\rm SI} \lesssim (7-9) \times 10^{-46} {\rm cm}^2
\eeq
The SI-elastic scattering cross section for VDM to scatter off a proton target is given by
\bea
\sigma_p^{\rm SI} \label{sp-th}
&=& \frac{4 \mu_V^2}{\pi} \left(\frac{g_X s_\alpha c_\alpha m_p}{2 v_H}\right)^2 \left(\frac{1}{m_1^2}-\frac{1}{m_2^2}\right)^2 f_p^2,
\nonumber \\
&\simeq& 2.2 \times 10^{-45} {\rm cm}^2 \l( \frac{g_X s_\alpha c_\alpha}{10^{-2}} \r)^2 \l( \frac{75 \GeV}{m_\phi} \r)^4 \l( 1 - \frac{m_\phi^2}{m_h^2} \r)^2
\eea
where $\mu_V =m_V m_p/(m_V + m_p)$ and $f_p = 0.326$~\cite{Young:2009zb}  (see Ref.~\cite{Crivellin:2013ipa} for more recent analysis) was used.
Note that $m_\phi \sim m_h$ results in some amount of cancellation between contributions of $\phi$ and $h$ to $\sigma_p^{\rm SI}$. 
As the result, the LUX bound can be satisfied rather easily for 
$g_X s_\alpha c_\alpha \lesssim 10^{-2}$.  
As shown in Fig.~\ref{fig:alpha-gx-LUXbnd}, direct detection experiments leave a wide range of parameter space uncovered.
This is unfortunate since it implies that the model cannot be entirely cross checked by other physical observables.

\subsection{Dark matter relic density} \label{subsec:relic}

The observed $\GeV$ scale $\gamma$-ray spectrum may be explained if DM annihilates mainly into $b \overline{b}$ with a velocity-averaged annihilation cross section close to the canonical value of thermal relic dark matter.
This implies that $30 \GeV \lesssim m_V \lesssim 40 \GeV$ in case of the $s$-channel 
annihilation (Fig.~\ref{fig:dm-to-bb}) scenario.
It is also possible to produce $b\bar{b}$ with the nearly same energy from the decay of highly non-relativistic $\phi$ which is produced from the annihilation of DM having mass of $60 \GeV \lesssim m_V \lesssim 80 \GeV$ (Fig.~\ref{fig:dm-to-2phi}).
In both cases,  it is expected to have $\tau\bar{\tau}$ and $c\bar{c}$ productions too in the final states, because $H_1$ will decay into them with branching ratios about 7\% and 3\%.
\tikzset{
particle/.style={thick,draw=black, postaction={decorate},
    decoration={markings,mark=at position .5 with {\arrow{triangle 45}}}},
fermion/.style={thick,draw=black, postaction={decorate},
    decoration={markings,mark=at position .5 with {\arrow{triangle 45}}}},
scalar/.style={thick,dashed,draw=black},
photon/.style={decorate, draw=black,
    decoration={coil,aspect=0}}
}
\begin{figure}
\centering
\begin{tikzpicture}[node distance=1.5cm and 1cm]
\coordinate[label=left:$V^{\mu}$] (e1);
\coordinate[below right=of e1] (aux1);
\coordinate[below left=of aux1,label=left:$V^{\nu}$] (e3);
\coordinate[right=1.25cm of aux1] (aux2);
\coordinate[above right=of aux2,label=right:$\bar{b} / \bar{\tau}$] (e2);
\coordinate[below right=of aux2,label=right:$b / \tau$] (e4);

\draw[photon] (e1) -- (aux1);
\draw[photon] (e3) -- (aux1);
\draw[fermion] (aux2) -- (e2);
\draw[fermion] (e4) -- (aux2);
\draw[scalar] (aux1) -- node[label=above:$H_{1,2}$] {} (aux2);
\end{tikzpicture}
\caption{\label{fig:dm-to-bb} Dominant $s$ channel $b+\bar{b}$ (and $\tau+\bar{\tau}$) production}
\end{figure}
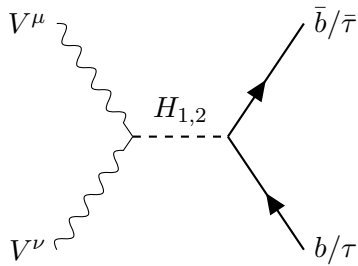
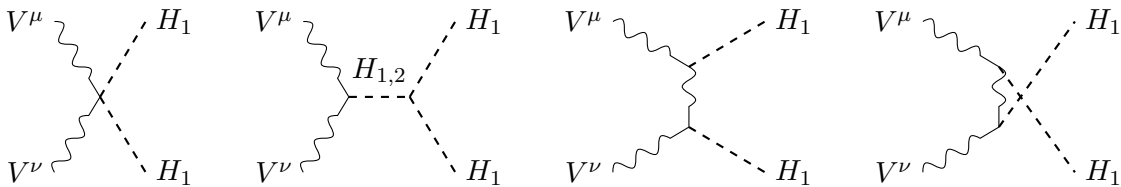
\begin{figure}
\centering
\begin{tikzpicture}[node distance=1.0cm and 0.6cm]
\coordinate[label=left:$V^{\mu}$] (e1);
\coordinate[below right=of e1] (aux1);
\coordinate[below left=of aux1,label=left:$V^{\nu}$] (e3);
\coordinate[above right=of aux1,label=right:$H_1$] (e2);
\coordinate[below right=of aux1,label=right:$H_1$] (e4);

\draw[photon] (e1) -- (aux1);
\draw[photon] (e3) -- (aux1);
\draw[scalar] (aux1) -- (e2);
\draw[scalar] (aux1) -- (e4);
\end{tikzpicture}
\hspace{3mm}
\begin{tikzpicture}[node distance=1.0cm and 0.6cm]
\coordinate[label=left:$V^{\mu}$] (e1);
\coordinate[below right=of e1] (aux1);
\coordinate[below left=of aux1,label=left:$V^{\nu}$] (e3);
\coordinate[right=0.8cm of aux1] (aux2);
\coordinate[above right=of aux2,label=right:$H_1$] (e2);
\coordinate[below right=of aux2,label=right:$H_1$] (e4);

\draw[photon] (e1) -- (aux1);
\draw[photon] (e3) -- (aux1);
\draw[scalar] (aux2) -- (e2);
\draw[scalar] (aux2) -- (e4);
\draw[scalar] (aux1) -- node[label=above:$H_{1,2}$] {} (aux2);
\end{tikzpicture}
\hspace{3mm}
\begin{tikzpicture}[node distance=0.6cm and 1.0cm]
\coordinate[label=left:$V^{\mu}$] (e1);
\coordinate[below right=of e1] (aux1);
\coordinate[above right=of aux1,label=right:$H_1$] (e2);
\coordinate[below=0.8cm of aux1] (aux2);
\coordinate[below left=of aux2,label=left:$V^{\nu}$] (e3);
\coordinate[below right=of aux2,label=right:$H_1$] (e4);

\draw[photon] (e1) -- (aux1);
\draw[scalar] (aux1) -- (e2);
\draw[photon] (e3) -- (aux2);
\draw[scalar] (aux2) -- (e4);
\draw[photon] (aux1) --  node[] {} (aux2);
\end{tikzpicture}
\hspace{3mm}
\begin{tikzpicture}[node distance=0.6cm and 1.0cm]
\coordinate[label=left:$V^{\mu}$] (e1);
\coordinate[below right=of e1] (aux1);
\coordinate[above right=of aux1,label=right:$H_1$] (e2);
\coordinate[below=0.8cm of aux1] (aux2);
\coordinate[below left=of aux2,label=left:$V^{\nu}$] (e3);
\coordinate[below right=of aux2,label=right:$H_1$] (e4);

\draw[photon] (e1) -- (aux1);
\draw[scalar] (aux1) -- (e4);
\draw[photon] (e3) -- (aux2);
\draw[scalar] (aux2) -- (e2);
\draw[photon] (aux1) -- node[] {} (aux2);
\end{tikzpicture}
\caption{\label{fig:dm-to-2phi} Dominant $s/t$-channel production of $H_1$s that decay dominantly to $b+\bar{b}$}
\end{figure}

In the process of Fig.~\ref{fig:dm-to-bb}, the thermally-averaged annihilation cross section 
of VDM is given by 
\beq \label{sv-ff}
\langle \sigma v_{\rm rel} \rangle_{f\bar{f}}
= \sum_f \frac{\l( g_X s_\alpha c_\alpha \r)^2}{3 \pi} m_X^2 \l| \sum_i \frac{1}{s-m_i^2 + 
i m_i \Gamma_i} \r|^2 \l( \frac{m_f}{v_H} \r)^2 \l( 1-\frac{4 m_f^2}{s} \r)^{3/2}   ,
\eeq
where $m_f$ is the mass of a SM fermion $f$.
Note that \eq{sv-ff} is suppressed by a factor $s_\alpha^2 m_f^2$.
Hence a large enough annihilation cross section for the right amount of relic density can be achieved only around the resonance region.
However in the resonance region the annihilation cross section varies a lot, as the Mandalstam $s$-variable varies from the value at freeze-out to the value in a dark matter halo at present.
Therefore, this process can not be used for the $\GeV$ scale $\gamma$-ray spectrum from the galactic center.

On the other hand, in the process of Fig.~\ref{fig:dm-to-2phi} for 
$m_\phi < m_V \lesssim 80 \GeV$, the thermally-averaged annihilation cross section 
of VDM is given by 
\beq
\langle \sigma v_{\rm rel} \rangle_{\rm tot} = \langle \sigma v_{\rm rel} \rangle_{f\bar{f}} + \langle \sigma v_{\rm rel} \rangle_{\phi\phi}
\eeq
where 
\bea
\langle \sigma v_{\rm rel} \rangle_{\phi\phi}
&\simeq& \frac{1}{16 \pi s} \overline{\l| \mathcal{M} \r|^2} \l( 1 - \frac{4 m_\phi^2}{s} \r)^{1/2}
\eea
with 
\bea
\overline{\l| \mathcal{M} \r|^2}
&\approx& \frac{2}{9} \l[ 1 + 4 \l( \frac{s}{4 m_V^2} \r)^2 \l( 1 - \frac{2 m_V^2}{s} \r)^2 \r] \l[ \l( 2 c_\alpha^2 g_X^2 + \mathcal{M}_s^0 \r) - 8 c_\alpha^2 g_X^2 \r]^2
\\
\mathcal{M}_s^0 
&=& 2 c_\alpha^4 m_V^2 \l( \frac{6 \lambda_\Phi}{s - m_\phi^2} - \frac{t_\alpha \lambda_{\Phi H} v_H/v_\Phi}{s - m_h^2} \r)
\simeq 4 c_\alpha^4 \lambda_\Phi \l[ 1 - \frac{s_\alpha^2 m_V^2 \l( m_h^2 - m_\phi^2 \r)}{m_\phi^2 \l( s - m_h^2 \r)} \r]
\nonumber \\
&\sim& 2 c_\alpha^4 g_X^2 \l[ 1 - \frac{s_\alpha^2 \l( m_h^2 - m_V^2 \r)}{\l( 4 m_V^2 - m_h^2 \r)} \r]
\eea
Note that, if we consider the off-resonance region with $2 m_V \nsim m_h$, 
the contribution of the $s$-channel $H_2$ mediation can be ignored and 
$\langle \sigma v_{\rm rel} \rangle_{\phi\phi}$ does not depend neither 
$s_\alpha$ nor $m_f$.   Hence a right size of annihilation cross section can be obtained 
by adjusting mostly $g_X$ and $\l( m_V - m_\phi \r)/m_V$, with the negligible mixing angle 
dependence.  
%
\begin{figure}[h]
\centering
\includegraphics[width=0.7\textwidth]{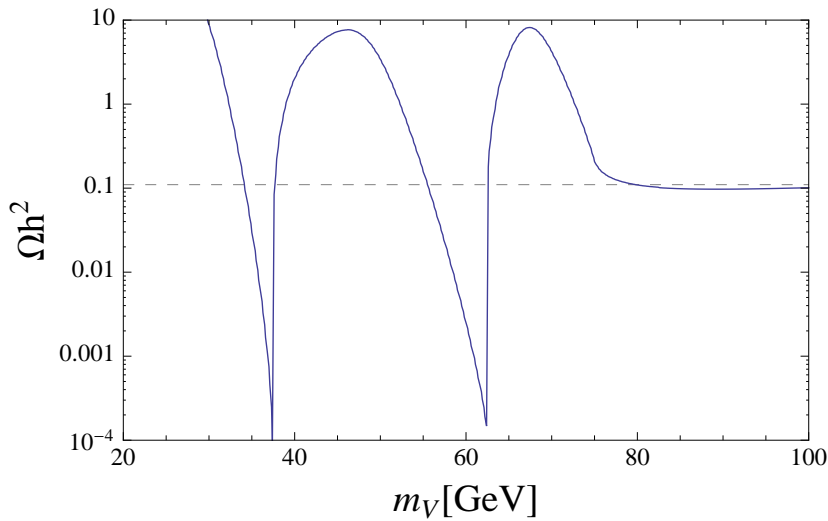}
\caption{\label{fig:OmegaDM} 
Relic density of dark matter as function of $m_\psi$ for $m_h=125$, $m_\phi = 75 \GeV$, $g_X = 0.2$, and $\alpha=0.1$.
}
\end{figure}
Fig.~\ref{fig:OmegaDM} shows the relic density at present 
\footnote{We adapted the micrOMEGAs package~\cite{Belanger:2008sj,micromegas} 
($\Omega_{\rm VDM} h^2$)  to our model for numerical calculation.} as a function of 
$m_V$  for $m_\phi = 75$ GeV and $g_X = 0.2$ and the mixing angle   $\alpha = 0.1$.
From Fig.~\ref{fig:OmegaDM}, we note that the mass of our VDM is constrained to be 
$m_h/2 < m_V$, since SM-Higgs resonance should be also avoided. 
And the velocity-averaged annihilation cross section at present epoch can be close 
to that of freeze-out only for $m_\phi \lesssim m_V$.
\begin{figure}
\centering
\includegraphics[scale=0.8]{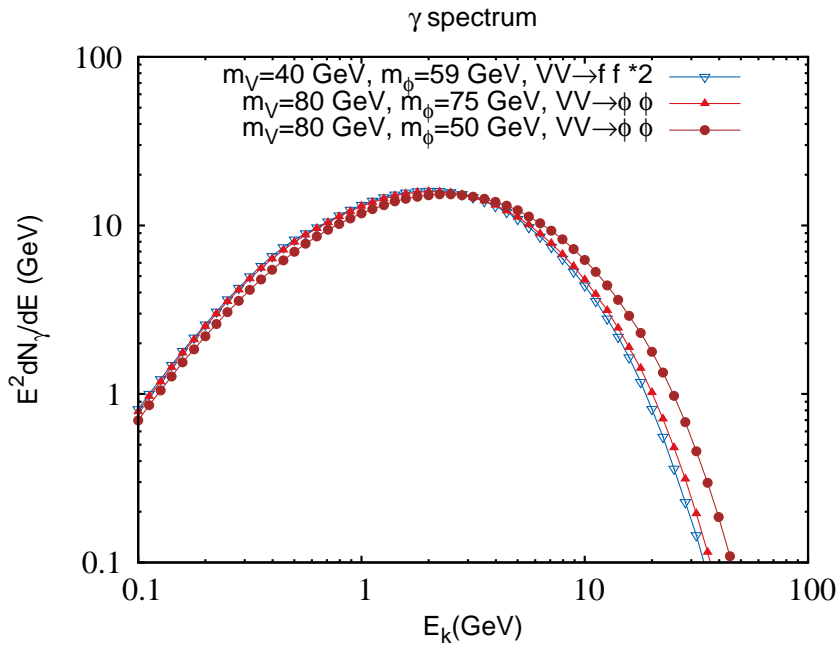}
\caption{Illustration of $\gamma$ spectra from different channels. The first two cases give almost the same spectra while in the third case $\gamma$ is boosted so the spectrum is shifted to higher energy.\label{fig:gammaspectra}}
\end{figure}
Note also that, as shown in Fig.~\ref{fig:gammaspectra}, in order to match to the observed $\gamma$-ray spectrum, we need $m_\phi \sim m_V$ to avoid boosted $\phi$.

In the region of $60 \GeV \lesssim m_\phi \sim m_V \lesssim 80 \GeV$, the SM Higgs 
boson decay into VDM is suppressed by the phase space factor or kinematically forbidden.
Hence the collider bound on the invisible decay of SM Higgs is irrelevant, but the mixing 
angle is still constrained by the signal strength of SM channels such that 
$\alpha \lesssim 0.4$ \cite{Chpoi:2013wga}.

A remark is in order for the present annihilation cross section to obtain observed 
$\GeV$ scale $\gamma$-ray.   
Compared to the case of $30 \GeV \lesssim m_V \lesssim 40 \GeV$, the present number 
density of dark matter for $60 \GeV \lesssim m_V \lesssim 80 \GeV$ is smaller 
by a factor of about a half, but each annihilation produces two pairs of $b\bar{b}$.
Hence, the expected flux which is proportional to the square of DM number density is 
smaller by about a half.  However, there are various astrophysical uncertainties 
in the estimation of required annihilation cross section.  
In particular, a small change of the inner slope of DM density profile is enough 
to compensate the difference of about factor two.   In addition, as discussed in  
Refs.~\cite{Daylan:2014rsa}, the $\GeV$ scale $\gamma$-ray data fits well to cross 
sections proportional to the square of the mass of the final state SM particles.
This kind of flavor-dependence is an intrinsic nature of our SVDM scenario,   
thanks to the Higgs portal interaction.   Therefore, with these points in mind, 
SVDM with mass of $60 \GeV \lesssim m_V \lesssim 80 \GeV$ can be a natural source 
of the $\GeV$ scale $\gamma$-ray excess from the direction of the galactic center.

\subsection{Comparison with other Higgs portal DM models}
In regard of the $\GeV$ scale $\gamma$-ray excess from the galactic center, 
SSDM can work equally well as our SVDM scenario.  
One difference from SVDM is the additional Higgs portal interaction of SSDM 
with SM Higgs, which can improve the vacuum instability problem of SM Higgs 
potential better than SVDM scenario.

Contrary to SSDM or SVDM, SFDM with a real scalar mediator results in $p$-wave 
$s$-channel annihilation.  In addition, the $t$-channel annihilation cross section 
is approximately proportional to $\l( 1 - m_\phi^2/m_{\rm DM}^2 \r)^{3/2}$ in the 
low momentum limit.   Since $\l( m_{\rm DM}  - m_\phi \r)/m_{\rm DM} \ll 1$ is 
needed in order to avoid a boosted $\phi$, such a $t$-channel annihilation in 
SFDM scenario is suppressed by an additional factor $\l( 1 - m_\phi^2/m_{\rm DM}^2 \r)$ relative to the case of SSDM and SVDM.   Hence SFDM needs a pseudo-scalar mediator 
and it makes model a bit complicated (see for example Ref.~\cite{Ipek:2014gua}).

Contrary to the case of SFDM where a wide range of pseudo-scalar mass is allowed, 
the requirement of the $t$-channel annihilation of DM near threshold in SSDM and SVDM  
constrains the mass of non-SM Higgs $\phi$ to be within a narrow range of 
\beq
m_h/2 \lesssim m_\phi \lesssim 80 \GeV
\eeq
Therefore, dedicated searches of the second Higgs boson at future collider experiments 
can focus on this range of invariant mass although too small mixing angle or too small 
trilinear coupling for $H-\phi-\phi$ would end up a null result.

\section{Conclusion} \label{sec:conc}
In this paper, we revisited the singlet vector dark matter (SVDM) model with 
Higgs portal in order to see if it can explain the observed $\GeV$ scale $\gamma$-ray 
spectrum from the galactic center by the annihilation of dark matter mainly to 
$b\bar{b}$ or to two non-SM light Higgses which decay subsequently and dominantly 
to $b\bar{b}$.  We found that the Higgs portal SVDM scenario can naturally explain 
the $\gamma$-ray spectrum while providing a right amount of relic density for 
$m_h/2 < m_V \lesssim 80 \GeV$ and $(m_V-m_\phi)/m_V \ll 1$ with $m_V$ and $m_\phi$ 
being the masses of VDM and non-SM Higgs boson. 
This implies that the mass of the non-SM extra Higgs is constrained 
to be within a narrow range of 
\beq
m_h/2 \lesssim m_\phi \lesssim 80 \GeV
\eeq
which can be focused on in dedicated searches of the second Higgs at future 
collider experiments although a null result due to very small mixing angle $\alpha$ 
is also possible.   The dark gauge coupling is contained to be $g_X \sim 0.2$ for 
the right amount of relic density while taking $\alpha$ to be small enough to satisfy 
direct DM search bound. Unfortuantely the LUX or XENON1T cannot explore the entire 
parameter space of the SVDM explaining the GeV-scale $\gamma$-ray from the galactic 
center.  The instability of SM vacuum could be improved due to the additional loop 
contribution of an extra scalar field.

\acknowledgments

This work is supported in part by National Research Foundation of Korea (NRF) Research Grant 2012R1A2A1A01006053 (PK, WIP, YT), and by SRC program of NRF funded by 
MEST (20120001176) through Korea Neutrino Research Center at Seoul National 
University (PK).

%


\end{document}